\documentclass[10pt, Conference, letterpaper]{IEEEtran}
\usepackage{graphicx}
\usepackage{amssymb}
\usepackage{subfigure}
\usepackage{amsthm}
\usepackage{amsmath}
\usepackage{caption}
\usepackage{float}
\usepackage{color}

\newtheorem{theorem}{Theorem}

\usepackage{dcolumn}
\usepackage{algorithm}
\usepackage{algorithmic}

\usepackage{bm}
\usepackage{mathtools, cuted}
\usepackage[english]{babel}
\usepackage{lipsum}
\usepackage{multicol}
\usepackage{mathtools, cuted}
\usepackage{booktabs}
\usepackage{multirow}
\makeatletter
\renewcommand{\maketag@@@}[1]{\hbox{\m@th\normalsize\normalfont#1}}%
\makeatother

\allowdisplaybreaks[3]
\usepackage{geometry}
\usepackage{geometry}
\begin{document}

\title{A Finite Blocklength Approach for Wireless Hierarchical Federated Learning in the Presence of Physical Layer Security}

\author{Haonan~Zhang, Chuanchuan~Yang, Bin~Dai}

\author{\IEEEauthorblockN{
Haonan~Zhang\IEEEauthorrefmark{1}\IEEEauthorrefmark{4},
Chuanchuan~Yang\IEEEauthorrefmark{3}\IEEEauthorrefmark{4},
Bin~Dai\IEEEauthorrefmark{1}\IEEEauthorrefmark{4}\\}
\IEEEauthorblockA{
\IEEEauthorrefmark{1}
School of Information Science and Technology, Southwest Jiaotong University, Chengdu, 610031, China.\\}
\IEEEauthorblockA{
\IEEEauthorrefmark{3}
Department of Electronics, Peking University, Beijing, 100871, China.\\}
\IEEEauthorblockA{
\IEEEauthorrefmark{4}
Peng Cheng Laboratory, Shenzhen, 518055, China.\\
zhanghaonan@my.swjtu.edu.cn, yangchuanchuan@pku.edu.cn, daibin@home.swjtu.edu.cn.}
}

\maketitle

\begin{abstract}
The wireless hierarchical federated learning (HFL) in the presence of physical layer security (PLS) issue is revisited.
Though a framework of this problem has been established in the previous work, practical secure finite blocklength (FBL) coding scheme remains unknown.
In this paper, we extend the already existing FBL coding scheme for the white Gaussian channel with noisy feedback to the wireless HFL with
quasi-static fading duplex channel, and derive achievable rate and upper bound on the eavesdropper's uncertainty of the extended scheme.
The results of this paper are further explained via simulation results.
\end{abstract}


\section{Introduction \label{secI}}
\setcounter{equation}{0}

\begin{figure}[htb]
\centering
\includegraphics[scale=0.25]{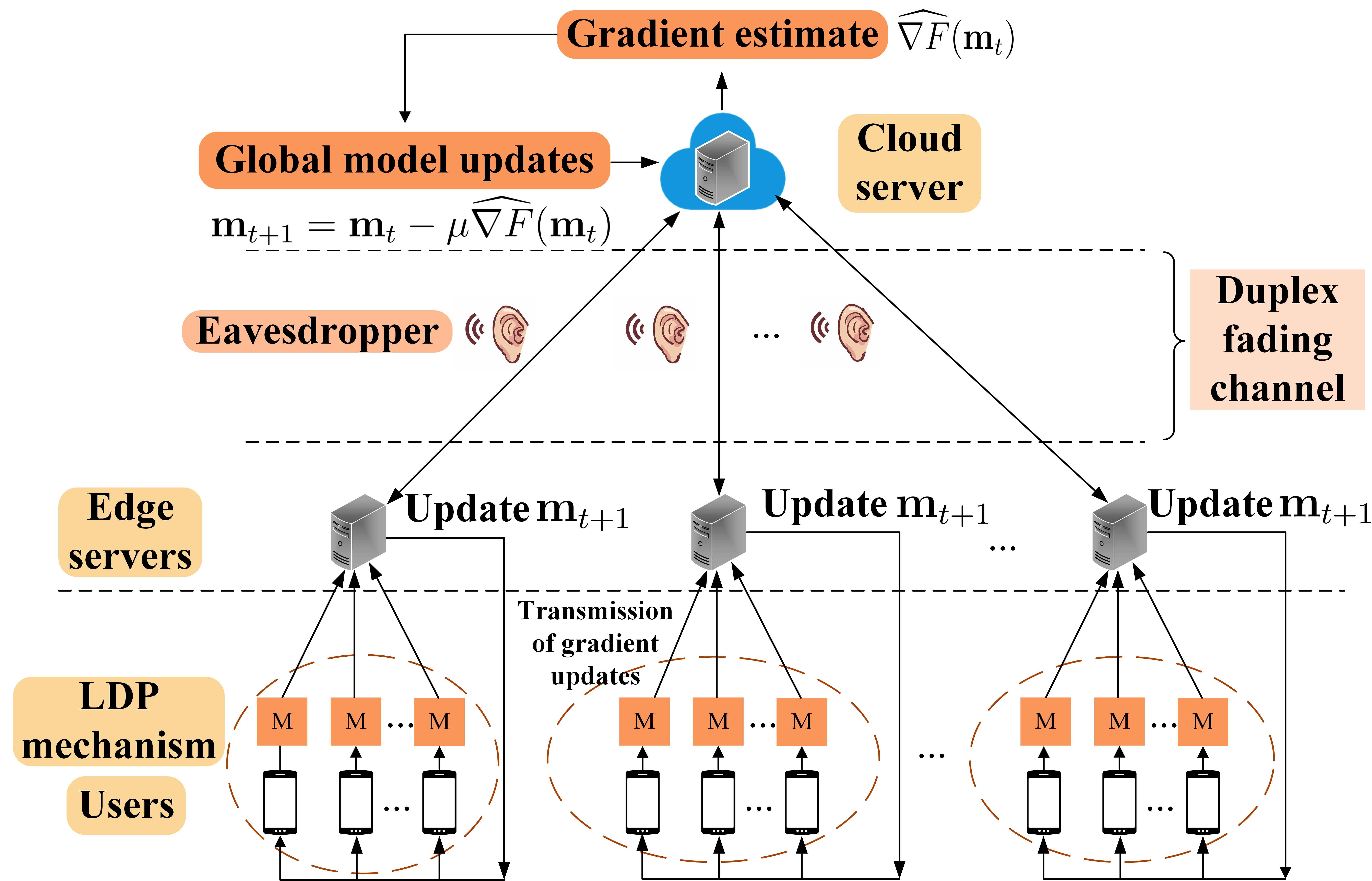}
\caption{The wireless HFL in the presence of eavesdroppers}
\label{fig1}
\end{figure}

The wireless federated learning has been extensively studied in the literature \cite{GZB}-\cite{b}. Recently,
with the development of edge computing, the client-edge-cloud hierarchical federated learning (HFL) systems receive much attention \cite{SLHFL1}-\cite{LL}.
However, due to the broadcast nature of wireless communications, the wireless FL is susceptible to eavesdropping.
In this paper, we study the wireless HFL in the presence of eavesdropping, see Figure \ref{fig1}. In Figure \ref{fig1},
users, edge servers and the cloud server cooperate with each other to jointly train a learning model, and in the meanwhile,
the malicious cloud server may
infer the presence of an individual data sample from a learnt
model by various attacks. Differential privacy (DP)  has been proved to be an effective way to protect the individual data against such attacks, and hence before aggregation of all users' gradients to the edge servers, the Gaussian noise which is used as local differential privacy (LDP) mechanism \cite{dp2} is added to the gradient of each user. Moreover, each edge server communicates with the cloud server via a duplex fading channel, and due to the broadcast nature of wireless communication, this channel is eavesdropped by an external eavesdropper. The main object of Figure \ref{fig1}
is to \emph{minimize the information leakage to the malicious cloud server subject to a certain mount of utility of the polluted data gradients (added by the Gaussian noises), and protect the transmitted data in wireless channels from eavesdropping}. In \cite{infocom}, the fundamental limit in the utility-privacy-physical layer security (PLS) trade-off was established. However, note that the secrecy capacity in \cite{infocom} cannot be approached by
finite blocklength (FBL) coding scheme since it is proved by using random binning coding scheme \cite{Wy}.
Then it is natural to ask: can we design a constructive FBL coding scheme
for the edge server, and confuse the eavesdropper as much as possible?

In this paper, first, we extend an already existing FBL coding scheme for the white Gaussian channel with noisy feedback \cite{feedback} to the
model of Figure \ref{fig1}, then we derive achievable rate and upper bound on the eavesdropper's uncertainty of the extended scheme.
Finally, we show the relationship between utility, privacy, PLS and other parameters via simulation examples.

\section{ Preliminary, Model Formulation and Main Results}\label{secII}
\setcounter{equation}{0}

\subsection{Preliminary: learning protocol}\label{secII-2}

In Figure \ref{fig1}, there are a cloud server, $L$ edge servers indexed by $\ell$, and $K$ users indexed by $k$ and $\ell$. $\{\mathcal{C}_{\ell}\}_{\ell=1}^{L}$ represents the disjoint user sets and $|\mathcal{C}_{\ell}|$ is the number of users in edge domain $\ell$, $\{\mathcal{S}_{\ell,k}\}_{k=1}^{|\mathcal{C}_{\ell}|}$ represents the distributed datasets and $S_{\ell,k}=|\mathcal{S}_{\ell,k}|$ is the cardinality of $\mathcal{S}_{\ell,k}$, where $\mathcal{S}_{\ell,k}= \{(\mathbf{u}_{k,j},v_{k,j})\}_{j=1}^{|\mathcal{S}_{\ell,k}|}$, $\mathbf{u}_{k,j}\in \mathbb{R}^{q}$ is the $j$-th vector of covariates with $q$ features and $v_{k,j}\in \mathbb{R}$ is the corresponding associated label at user $k$. Denote the aggregated dataset in edge $\ell$ domain by $\mathcal{S}_{\ell}$, and each edge server aggregates gradients from its users. The global loss function $F(\mathbf{m})$ is given by
\begin{small}
\begin{align}\label{gloss}
F(\mathbf{m})=\frac{1}{S}\sum_{\ell=1}^{L}\sum_{k=1}^{|\mathcal{C}_{\ell}|}S_{\ell,k}F_{\ell,k}(\mathbf{m}),
\end{align}
\end{small}where $\mathbf{m}\in \mathbb{R}^{q}$ is the model vector and $S=\sum_{\ell}\sum_{k}S_{\ell,k}$. $F_{\ell,k}(\cdot)$ is the local loss function for user $k$, where
\begin{small}
\begin{align}\label{lloss}
F_{\ell,k}(\mathbf{m})=\frac{1}{S_{\ell,k}}\sum_{(\mathbf{u}_{k,j},v_{k,j})\in\mathcal{S}_{\ell,k}}f(\mathbf{m};\mathbf{u}_{k,j},v_{k,j})+\lambda R(\mathbf{m}),
\end{align}
\end{small}and $f(\mathbf{m};\mathbf{u}_{k,j},v_{k,j})$ is the sample-wise loss function. $R(\mathbf{m})$ is a strongly convex regularization function and $\lambda\geq0$. The model training by minimizing the global loss function as
\begin{small}
\begin{align}\label{mloss}
\mathbf{m}^{\star}=\arg\min_{\mathbf{m}}F(\mathbf{m}).
\end{align}
\end{small}To minimize $F(\mathbf{m})$, we use a distributed gradient descent iterative algorithm.
Specifically, \emph{in the $t$-th ($t\in\{1,2,...,T\}$) communication round (the overall communication round is $T$)}, each user $k$ computes its own local gradient
$\nabla F_{\ell,k}(\mathbf{m}_{t})$
and the users send the corrupted local gradients (added by Gaussian noises for LDP) to the edge servers. Then, the edge server $\ell$ computes its estimation $\widehat{ \nabla F_{\ell}}(\mathbf{m}_{t})$ of the partial gradient and $\nabla F_{\ell}(\mathbf{m}_{t})=\frac{1}{S_{\ell}}\sum_{k\in\mathcal{C}^{\ell}}S_{\ell,k}\nabla F_{\ell,k}(\mathbf{m}_{t})$,
where $S_{\ell}=|\mathcal{S}_{\ell}|$ is the total number of $\mathcal{S}_{\ell}$. The cloud server's estimation $\widehat{ \nabla F}(\mathbf{m}_{t})$ of the global gradient is given by
$\nabla F(\mathbf{m}_{t})=\frac{1}{S}\sum_{\ell=1}^{L}S_{\ell}\nabla F_{\ell}(\mathbf{m}_{t})$.
The global model $\mathbf{m}_{t+1}$ updated by the cloud server is given by
\begin{small}
\begin{align}\label{update}
\mathbf{m}_{t+1}=\mathbf{m}_{t}-\mu\widehat{ \nabla F}(\mathbf{m}_{t}),
\end{align}
\end{small}where $\mu$ is the learning rate. For convenience, in the $t$-th communication round, we denote $\mathbf{W}_{t,k}=S_{\ell,k}\nabla F_{\ell,k}(\mathbf{m}_{t})$.

\subsection{Model formulation }\label{secII-3}

\begin{figure}[htb]
\centering
\subfigure[An information-theoretic approach of Figure \ref{fig1}: encoding]{
\includegraphics[scale=0.14]{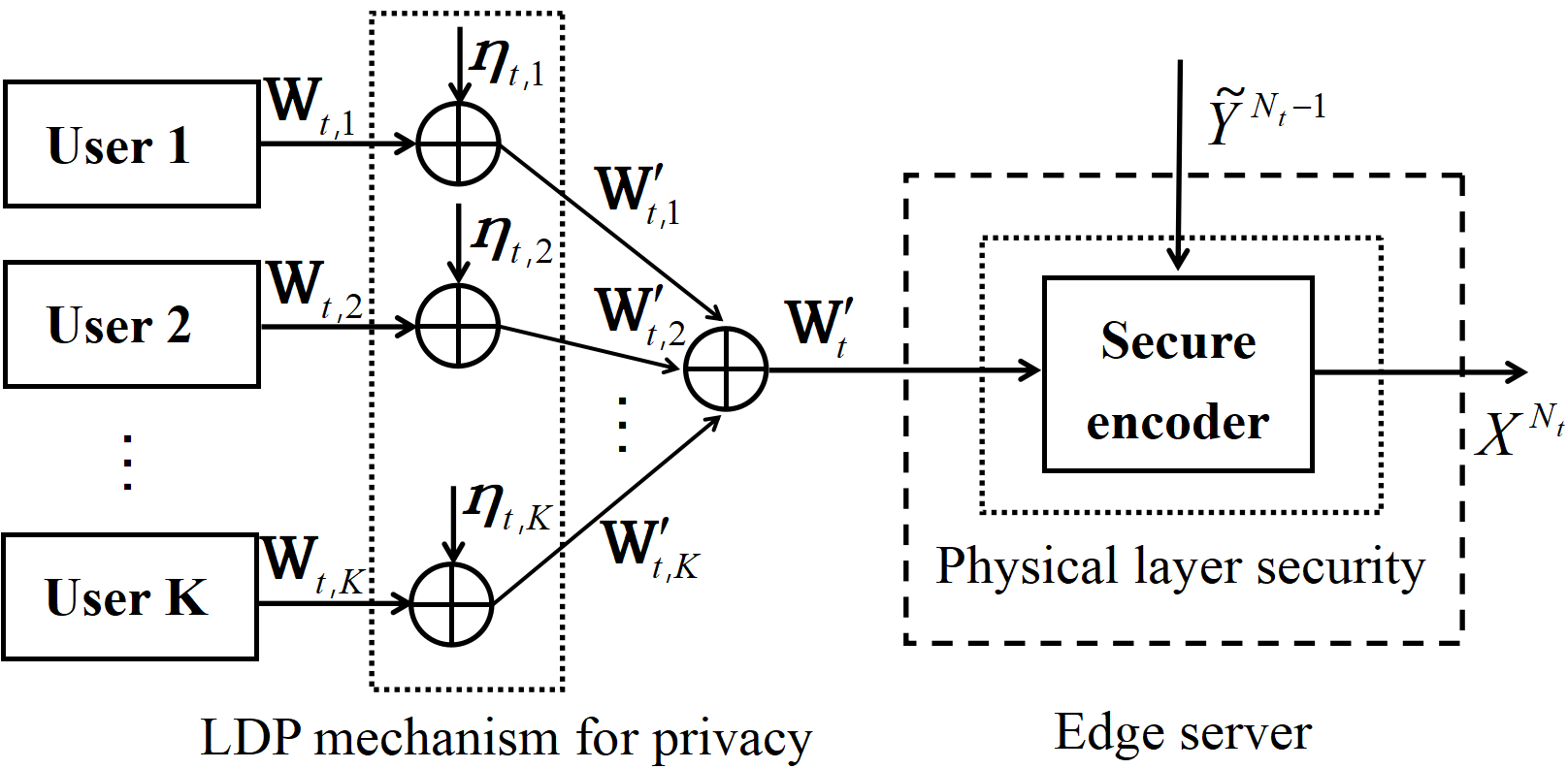}\label{fig2-1}}\\
\subfigure[An information-theoretic approach of Figure \ref{fig1}: decoding]{
\includegraphics[scale=0.139]{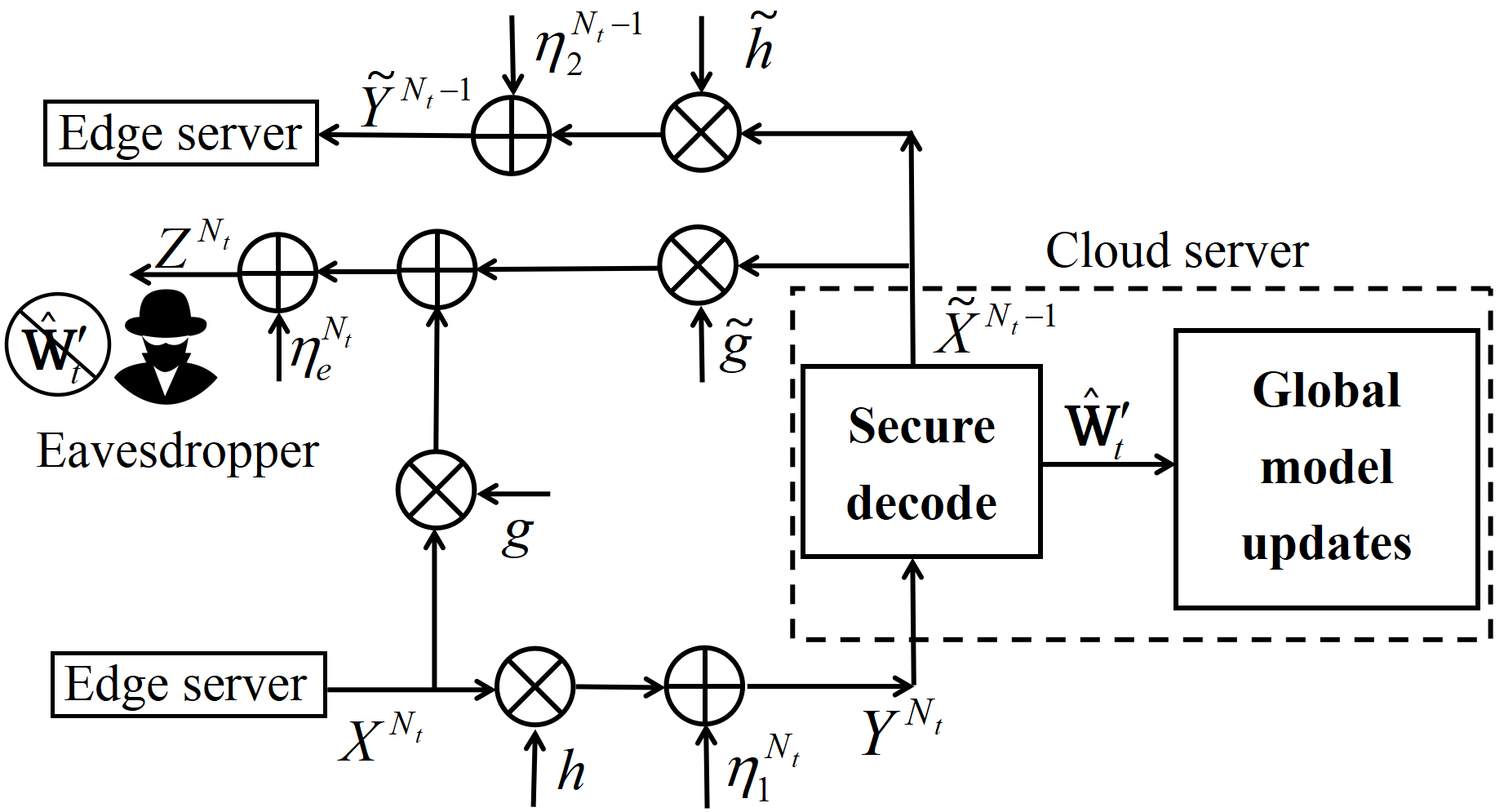}\label{fig2-2}}
\caption{An information-theoretic approach of Figure \ref{fig1}}
\label{fig2}
\end{figure}

In this paper, we assume that each edge server communicates with the cloud server without interference from other edge servers. Besides, we assume that the downlink communication is perfect, which is similar to \cite{DL}, and eavesdropper only shows interest in the data transmitted in the uplink communication between the edge servers and the cloud server. Hence we only focus on \textbf{the $T$ rounds uplink communication} between one of the edge servers and the cloud server. An information-theoretic approach of Figure \ref{fig1}
is illustrated in Figure \ref{fig2}.
For simplification, we make the following assumptions:
\begin{itemize}
  \item Similar to \cite{DL}-\cite{MS}, we assume that
the channel coefficients stay constants during the transmission (quasi-static fading channel).

  \item Similar to \cite{MS}-\cite{b}, we assume that the cloud server and the edge server have perfect channel state information (CSI) of the feedforward channel and feedback channel.

  \item From similar arguments in \cite{AM}, we assume that eavesdropper is an active user but it is un-trusted by the cloud server, which indicates that the perfect CSI of eavesdropper's channel is known by the eavesdropper and the edge server. Moreover, we assume that the eavesdropper also knows the perfect CSI of the edge server-cloud server's channels.
\end{itemize}

\emph{Information source}: In Figure \ref{fig2-1}, we assume that $\mathbf{W}_{t,k}\in \mathbb{R}^{q}$ is the $k$-th ($k\in\{1,2,...,K\}$) user's overall local gradient vector in $t$-th ($t\in\{1,2,...,T\}$) communication round, where $\mathbf{W}_{t,k}=(W_{t,k,1},...,W_{t,k,q})^{\mathcal{T}}$. Similar to \cite{ZJ}, the elements of $\mathbf{W}_{t,k}$ are independent and identically distributed (i.i.d.)
and $\mathbf{W}_{t,k}\sim \mathcal{N}(0,S_{\ell,k}\sigma_{w,t}^{2}\mathbf{I})$. Let $\boldsymbol\eta_{t,k}=(\eta_{t,k,1},...,\eta_{t,k,q})^{\mathcal{T}}$ be local artificial Gaussian noise i.i.d. according to distribution
$ \mathcal{N}(0,\sigma^{2}\mathbf{I})$. The corrupted local gradient $\mathbf{W}^{\prime}_{t,k}=(W^{\prime}_{t,k,1},...,W^{\prime}_{t,k,q})^{\mathcal{T}}$ that is aggregated by the edge server is given by
\begin{footnotesize}
\begin{eqnarray}\label{si}
&&\mathbf{W}^{\prime}_{t,k}=\mathbf{W}_{t,k}+\boldsymbol\eta_{t,k},
\end{eqnarray}
\end{footnotesize}%
where $\mathbf{W}^{\prime}_{t,k}\sim \mathcal{N}(0,(S_{\ell,k}\sigma_{w,t}^{2}+\sigma^{2})\mathbf{I})$ for $k\in\{1,2,...,K\}$. The overall local gradients and the overall noises are defined as  $\mathbf{W}_{t}=(W_{t,1},...,W_{t,q})^{\mathcal{T}}$ and $\boldsymbol\eta_{t}=(\eta_{t,1},...,\eta_{t,q})^{\mathcal{T}}$, respectively, where $W_{t,i}=\sum_{k=1}^{K}W_{t,k,i}$, $\eta_{t,i}=\sum_{k=1}^{K}\eta_{t,k,i}$ ($i\in\{1,2,...,q\}$). According to (\ref{si}), we define the overall corrupted local gradients sent to the edge server as $\mathbf{W}^{\prime}_{t}=(W^{\prime}_{t,1},...,W^{\prime}_{t,q})^{\mathcal{T}}$, where $W^{\prime}_{t,i}=\sum_{k=1}^{K}W^{\prime}_{t,k,i}$ ($i\in\{1,2,...,q\}$). Here note that since $\mathbf{W}_{t,k}$ and $\boldsymbol\eta_{t,k}$ are i.i.d. generated,
$\mathbf{W}^{\prime}_{t}$ is also composed of i.i.d. components, where $\mathbf{W}^{\prime}_{t}\sim \mathcal{N}(0,(S_{\ell}\sigma_{w,t}^{2}+K\sigma^{2})\mathbf{I})$.

\textbf{Definition 1} (Privacy by mutual information \cite{WW}): If the mutual information between $\mathbf{W}_{t}$ and $\mathbf{W}^{\prime}_{t}$ satisfies $\frac{1}{qT}\sum_{t=1}^{T}I(\mathbf{W}_{t};\mathbf{W}^{\prime}_{t}) \leq \epsilon$, we say the LDP mechanism satisfies $ \epsilon $-mutual-information privacy for some $\epsilon > 0 $.

\textbf{Definition 2} (Utility by quadratic distortion \cite{network}): The utility of $\mathbf{W}^{\prime}_{t}$ is characterized by
$d(\mathbf{W}_{t},\mathbf{W}^{\prime}_{t})=||\mathbf{W}^{\prime}_{t}-\mathbf{W}_{t}||^{2}$, where $||\textbf{X}||$ represents the $l_{2}$-norm of the vector $\textbf{X}$. If $\frac{1}{qT}\sum_{t=1}^{T}E(d(\mathbf{W}_{t},\mathbf{W}^{\prime}_{t}))\leq U$, we say the utility of $\mathbf{W}^{\prime}_{t}$ is up to $U$.

\emph{Channels}: At time instant $i$ ($i\in\{1,2,...,N_{t}\}$ of $t$-th  communication round, channel inputs and outputs are given by
\begin{footnotesize}
\begin{gather}
Y_{i}(t)=hX_{i}(t)+\eta_{1,i}(t),\ \ i=1,2,...,N_{t},\label{cio}\\
\widetilde{Y}_{i}(t)=\widetilde{h}\widetilde{X}_{i}(t)+\eta_{2,i}(t),\ \ i=1,2,...,N_{t}-1,\label{cio123} \\ Z_{i}(t)=gX_{i}(t)+\widetilde{g}\widetilde{X}_{i}(t)+\eta_{e,i}(t),\ \ i=1,2,...,N_{t},\label{cioe}
\end{gather}
\end{footnotesize}%
where $X_{i}(t)$ and $\widetilde{X}_{i}(t)$ respectively are the feedforward and feedback channel inputs, which satisfy the average power constraints $\frac{1}{N_{t}}\sum_{i=1}^{N_{t}}E[X_{i}(t)X_{i}(t)^{\mathcal{H}}]\leq P$ and  $\frac{1}{N_{t}-1}\sum_{i=1}^{N_{t}-1}E[\widetilde{X}_{i}(t)\widetilde{X}_{i}(t)^{\mathcal{H}}]\leq \widetilde{P}$. $h, \widetilde{h}, g, \widetilde{g}\in \mathbb{C}$ are the CSI of the feedforward and feedback channels of the cloud server, the feedforward and feedback channels of the eavesdropping channel, respectively, here note that $|h|$, $|\widetilde{h}|$, $|g|$ and $|\widetilde{g}|$ represent the modulus of $h$, $\widetilde{h}$, $g$ and $\widetilde{g}$.
$Y_{i}(t)$, $\widetilde{Y}_{i}(t)$ and $Z_{i}(t)$ respectively are the channel outputs of the cloud server, the edge server and the eavesdropper, $\eta_{1,i}(t)$, $\eta_{2,i}(t)$ and
$\eta_{e,i}(t)$ are i.i.d. as
$\mathcal{CN}(0,\sigma_{1}^{2})$, $\mathcal{CN}(0,\sigma_{2}^{2})$ and
$\mathcal{CN}(0,\sigma_{e}^{2})$, respectively. The signal-to-noise ratios of the feedforward and feedback channels are denoted by $\text{SNR}=\frac{P}{\sigma_{1}^{2}}$ and $\tilde{\text{SNR}}=\frac{\widetilde{P}}{\sigma_{2}^{2}}$.

\emph{Source coding}: We consider a lossy Gaussian source coding with quadratic distortion measure  $d(\mathbf{W}^{\prime}_{t},\hat{\mathbf{W}}^{\prime}_{t})=||\mathbf{W}^{\prime}_{t}-\hat{\mathbf{W}}^{\prime}_{t}||^{2}$, where $\hat{\mathbf{W}}^{\prime}_{t}$ is the estimation of source decoder. According to \cite[Chapter 3.8, pp. 64-65]{network}, there exists a source encoder mapping $ \mathbf{W}^{\prime}_{t}\rightarrow \{1,2,...,2^{qR_{t}(D)}\}$, it compresses $\mathbf{W}^{\prime}_{t}$  into an index $W^{\prime\prime}_{t}$ which is uniformly distributed in $\mathcal{W}^{\prime\prime}_{t} =\{1,2,...,2^{qR_{t}(D)}\}$, and the rate-distortion function $R_{t}(D)$ is given by
\begin{footnotesize}
\begin{align}\label{rate-distortion}
R_{t}(D)=\begin{cases}
\frac{1}{2}\log\frac{S_{\ell}\sigma_{w,t}^{2}+K\sigma^{2}}{D}& 0\leq D< S_{\ell}\sigma_{w,t}^{2}+K\sigma^{2}\\
0 & D\geq S_{\ell}\sigma_{w,t}^{2}+K\sigma^{2}
\end{cases}
\end{align}
\end{footnotesize}%
where $\frac{1}{qT}\sum_{t=1}^{T}E(d(\mathbf{W}^{\prime}_{t},\hat{\mathbf{W}}^{\prime}_{t}))\leq D$.
For the source decoder, a source decoding mapping maps $\{1,2,...,2^{qR_{t}(D)}\}$ to $\mathbf{\hat{W}}^{\prime}_{t}$.

\emph{Channel encoder}: At time $i$ ($i\in\{1,2,...,N_{t}\}$) in $t$-th ($t\in\{1,2,...,T\}$) communication round, the transmitted codeword $X_{i}(t)$ is a stochastic function of the $W^{\prime\prime}_{t}$, $h$, $\widetilde{h}$ and $\widetilde{Y}_{1}^{i-1}(t)=(\widetilde{Y}_{1}(t),...,\widetilde{Y}_{i-1}(t))$, i.e., $X_{i}(t)=f_{t,i}(W^{\prime\prime}_{t},h,\widetilde{h},\widetilde{Y}_{1}^{i-1}(t))$.

\emph{Channel decoder}: The channel decoder's estimation $\hat{w}^{\prime\prime}_{t}=\varphi(h,\widetilde{h},Y^{N_{t}})$, where $\varphi$ is the channel decoder's decoding function. The channel decoder with outputs $\widetilde{X}_{i}(t)=\widetilde{f}_{t,i}(h,\widetilde{h},Y_{1}^{i}(t))$, where $\widetilde{f}_{t,i}(\cdot)$ is a stochastic function.
The average decoding error probability of message $w^{\prime\prime}_{t}$ is given by
\begin{footnotesize}
\begin{align}\label{error1}
P_{e,t}=\frac{1}{|\mathcal{W}^{\prime\prime}_{t}|}\sum_{w^{\prime\prime}_{t}\in\mathcal{W}^{\prime\prime}_{t}}Pr\{\varphi(h,\widetilde{h},Y^{N_{t}})\neq w^{\prime\prime}_{t}|w^{\prime\prime}_{t}\,\, \mbox{sent}\}.
\end{align}
\end{footnotesize}%

\textbf{Definition 3}  The uncertainty of the eavesdropper (also called the secrecy level, which is first adopted in \cite{ET}) is defined as
 \begin{footnotesize}
\begin{equation}\label{eqfbl}
\Delta = \frac{H(W^{\prime\prime}_{1},...,W^{\prime\prime}_{T}|Z^{N_{1}},...,Z^{N_{T}},h,\widetilde{h},g,\widetilde{g})}{H(W^{\prime\prime}_{1},...,W^{\prime\prime}_{T})},\ \ 0\leq\Delta\leq1.
\end{equation}
\end{footnotesize}%
For fixed source encoding-decoding procedure, a rate $R$ is said to be $(N,\tau,U, \epsilon,\delta,D)$ achievable, if for given decoding error probability $\tau$, blocklength $N$, secrecy leve $\delta$, $\frac{1}{qT}\sum_{t=1}^{T}I(\mathbf{W}_{t};\mathbf{W}^{\prime}_{t}) \leq \epsilon$ and $\frac{1}{qT}\sum_{t=1}^{T}E(d(\mathbf{W}_{t},\mathbf{W}^{\prime}_{t}))\leq U$, there exists a channel code described above such that
\begin{footnotesize}
\begin{equation}\label{Def4}
\frac{H(W^{\prime\prime}_{1},...,W^{\prime\prime}_{T})}{N}=R,\,\,\, \frac{1}{T}\sum_{t=1}^{T}P_{e,t}\leq \tau,\,\,\ \Delta\geq\delta,
\end{equation}
\end{footnotesize}%
where $N=\sum_{t=1}^{T}N_{t}$.
The secrecy capacity $\mathcal{C}(N,\tau,U, \epsilon,\delta,D)$ is composed of all the secrecy achievable rates $R(N,\tau,U, \epsilon,\delta,D)$ defined above.
Here note that $\delta\in[0,1]$, and $\delta=1$ corresponds to perfect secrecy.
\subsection{Main result}\label{secII-4}
\begin{theorem}\label{Th1}
For given $N$, $\tau$, $U$, $\epsilon$, $\delta$ and $D$, a lower bound on the secrecy capacity $\mathcal{C}(N,\tau,U,\epsilon,\delta,D)$ is given by
\begin{footnotesize}
\begin{align}\label{Th1bound}
\mathcal{C}(N,\tau,U,\epsilon,\delta,D)\geq R(N,\tau,U,\epsilon,\delta,D),
\end{align}
\end{footnotesize}%
where
\begin{footnotesize}
 \begin{gather}
 R(N,\tau,U, \epsilon,\delta,D)=\frac{\sum_{t=1}^{T}N_{t}R_{t}}{N},\ \ N=\sum_{t=1}^{T}N_{t}\label{Th1R}\\
 R_{t}=
 \frac{1}{N_{t}}\log\left(\frac{3 \text{SNR}|h|^{2}}{\left[Q^{-1}(\frac{\tau}{8})\right]^{2}} \left(1+\frac{\text{SNR} |h|^{2}}{\Psi_{1}\Psi_{2}}\right)^{N_{t}-1}\right),\label{Rtotal2}\\
 \Psi_{1}=1+L\frac{|h|^{2}\text{SNR}}{|\widetilde{h}|^{2}\tilde{\text{SNR}}} ,\ \ \Psi_{2}=\left(1-\frac{L}{|\widetilde{h}|^{2}\tilde{\text{SNR}}}\right)^{-1},\label{parameters2-1}\\
 L=\frac{1}{3}\left[Q^{-1}(\frac{\tau}{8(N_{t}-1)})\right]^{2},\label{rng44}
 \end{gather}
 \end{footnotesize}%
$Q^{-1}(\cdot)$ is the inverse function of the Gaussian Q-function $Q(x)\stackrel{\text{def}}=\frac{1}{\sqrt{2\pi}}\int_{x}^{\infty}\exp(-\frac{u^{2}}{2})du$, and the uncertainty of the eavesdropper (the secrecy level) is upper bounded by
\begin{footnotesize}
\begin{equation}\label{Th1fblN}
\delta\leq \min_{t\in\{1,..,T\}}[1-\frac{\log\left(1+\frac{|g|^{2}P}{\sigma_{e}^{2}}\right)}{qR_{t}(D)}]^{+},\ \ [x]^{+}=\max(x,0),
\end{equation}
\end{footnotesize}%
where $R_{t}(D)$ is defined in (\ref{rate-distortion}), and
 \begin{footnotesize}
\begin{eqnarray}\label{Th1PU}
\max_{t\in\{1,..,T\}}\left\{\frac{S_{\ell}\sigma_{w,t}^{2}}{K(2^{2\epsilon}-1)}\right\}\leq\sigma^{2}\leq \frac{U}{K}.
\end{eqnarray}
\end{footnotesize}%

\end{theorem}
\begin{IEEEproof}
Theorem \ref{Th1} is proved by a FBL approach, which will be explained in the next section. The formal proof is also given in the next section.
\end{IEEEproof}

\section{A FBL Approach for wireless HFL}\label{secII-t}
\setcounter{equation}{0}
Since the FBL approach of each communication round is similar, we only describe the FBL approach of $t$-th ($t\in\{1,2,...,T\}$) communication round. Therefore, for simplify the notation, we omit the index $t$ of the signals and the noises in this section.
\subsection{Channel re-presentation}\label{secII-4-1}

At time $i$ ($i\in\{1,2,...,N_{t}\}$) in $t$-th communication round, since the elements in (\ref{cio})-(\ref{cio123}) are complex numbers, (\ref{cio})-(\ref{cio123}) can be re-written as
\begin{footnotesize}
\begin{align}\label{recio}
&Y_{R,i}+jY_{I,i}=(h_{R}+jh_{I})(X_{R,i}+jX_{I,i})+\eta_{R,1,i}+j\eta_{I,1,i},\nonumber\\
&\widetilde{Y}_{R,i}+j\widetilde{Y}_{I,i}=(\widetilde{h}_{R}+j\widetilde{h}_{I})(\widetilde{X}_{R,i}+j\widetilde{X}_{I,i})+\eta_{R,2,i}+j\eta_{I,2,i},
\end{align}
\end{footnotesize}%
where $j=\sqrt{-1}$, $Y_{R,i}=\text{Re}(Y_{i})$, $Y_{I,i}=\text{Im}(Y_{i})$, $h_{R}=\text{Re}(h)$, $h_{I}=\text{Im}(h)$,
$X_{R,i}=\text{Re}(X_{i})$, $X_{I,i}=\text{Im}(X_{i})$, $\eta_{R,1,i}=\text{Re}(\eta_{1,i})$, $\eta_{I,1,i}=\text{Im}(\eta_{1,i})$,
$\widetilde{Y}_{R,i}=\text{Re}(\widetilde{Y}_{i})$, $\widetilde{Y}_{I,i}=\text{Im}(\widetilde{Y}_{i})$, $\widetilde{h}_{R}=\text{Re}(\widetilde{h}),\widetilde{h}_{I}=\text{Im}(\widetilde{h})$,
$\widetilde{X}_{R,i}=\text{Re}(\widetilde{X}_{i})$, $\widetilde{X}_{I,i}=\text{Im}(\widetilde{X}_{i})$, $\eta_{R,2,i}=\text{Re}(\eta_{2,i})$, $\eta_{I,2,i}=\text{Im}(\eta_{2,i})$, where $\text{Re}(\cdot)$ and $\text{Im}(\cdot)$ denote the real and imaginary parts of a complex element, respectively. Here note that $E(X_{R,i}^{2})=P_{R}$, $E(X_{I,i}^{2})=P_{I}$, $E(\widetilde{X}_{R,i}^{2})=\widetilde{P}_{R}$ and $E(\widetilde{X}_{I,i}^{2})=\widetilde{P}_{I}$, where $P_{R}=P_{I}=\frac{1}{2}P$ and $\widetilde{P}_{R}=\widetilde{P}_{I}=\frac{1}{2}\widetilde{P}$.
From (\ref{recio}), we have
\begin{footnotesize}
\begin{align}\label{recio4}
X_{R(I),i}=Y^{\prime}_{R(I),i}-\eta^{\prime}_{R(I),1,i},\,\ \widetilde{X}_{R(I),i}=\widetilde{Y}^{\prime}_{R(I),i}-\eta^{\prime}_{R(I),2,i},
\end{align}
\end{footnotesize}%
where
$Y^{\prime}_{R,i}=(h_{R}Y_{R,i}+h_{I}Y_{I,i})/|h|^{2}$, $Y^{\prime}_{I,i}=(h_{R}Y_{I,i}-h_{I}Y_{R,i})/|h|^{2}$,
$\widetilde{Y}^{\prime}_{R,i}=(\widetilde{h}_{R}\widetilde{Y}_{R,i}+\widetilde{h}_{I}\widetilde{Y}_{I,i})/|\widetilde{h}|^{2}$, $\widetilde{Y}^{\prime}_{I,i}=(\widetilde{h}_{R}\widetilde{Y}_{I,i}-\widetilde{h}_{I}\widetilde{Y}_{R,i})/|\widetilde{h}|^{2},$
$\eta^{\prime}_{R,1,i}=(h_{R}\eta_{R,1,i}+h_{I}\eta_{I,1,i})/|h|^{2}$, $\eta^{\prime}_{I,1,i}=(h_{R}\eta_{I,1,i}-h_{I}\eta_{R,1,i})/|h|^{2}$,
$\eta^{\prime}_{R,2,i}=(\widetilde{h}_{R}\eta_{R,2,i}+\widetilde{h}_{I}\eta_{I,2,i})/|\widetilde{h}|^{2}$ and $\eta^{\prime}_{I,2,i}=(\widetilde{h}_{R}\eta_{I,2,i}-\widetilde{h}_{I}\eta_{R,2,i})/|\widetilde{h}|^{2}$.
Hence, (\ref{recio4}) is equivalent to (\ref{recio}), which indicates that the feedforward and feedback channels are divide into the two sub-channels. In addition, we conclude that $\text{Var}(\eta^{\prime}_{R,1,i})=\text{Var}(\eta^{\prime}_{I,1,i})=\sigma_{1}^{2}/2|h|^{2}$ and $\text{Var}(\eta^{\prime}_{R,2,i})=\text{Var}(\eta^{\prime}_{I,2,i})=\sigma_{2}^{2}/2|\widetilde{h}|^{2}$.

\subsection{Message splitting}\label{secII-4-22}
First, for given $D$, $N_{t}$, $\tau$, $\epsilon$ and $U$, let
 \begin{footnotesize}
\begin{align}\label{qR}
|\mathcal{W}^{\prime\prime}_{t}|=2^{N_{t}R_{t}}=2^{qR_{t}(D)},
\end{align}
\end{footnotesize}%
where $R_{t}=\frac{H(W^{\prime\prime}_{t})}{N_{t}}$. Here note that when $R_{t}(D)=0$, we do not transmit messages and choose $\mathbf{\hat{W}}^{\prime}_{t}=\bm{0}$. Then, the message $W^{\prime\prime}_{t}$ is divided into two independent parts $(W^{\prime\prime}_{t,R},W^{\prime\prime}_{t,I})$, where $W^{\prime\prime}_{t,R}$ and $W^{\prime\prime}_{t,I}$  respectively take values in $\mathcal{W}^{\prime\prime}_{t,R}=\{1,2,...,2^{N_{t}R_{t,R}}\}$ and $\mathcal{W}^{\prime\prime}_{t,I}=\{1,2,...,2^{N_{t}R_{t,I}}\}$, and $R_{t,R}+R_{t,I}=R_{t}$.
 Divide the interval $[-\sqrt{3},\sqrt{3}]$ into $2^{N_{t}R_{t,R}}(2^{N_{t}R_{t,I}})$ equally spaced sub-intervals, and the center of each sub-interval is mapped to a message value in $W^{\prime\prime}_{t,R}(W^{\prime\prime}_{t,I})$. Let $\theta_{R}(\theta_{I})$ be the center of the sub-interval with respect to (w.r.t) the message $W^{\prime\prime}_{t,R}(W^{\prime\prime}_{t,I})$, and $E(\theta_{R}^{2})=E(\theta_{I}^{2})=1$.
\subsection{Coding scheme}\label{secII-4-2}

For simplification, we only describe the coding scheme of message $W^{\prime\prime}_{t,R}$, and the coding scheme of message $W^{\prime\prime}_{t,I}$ is similar to the coding scheme of message $W^{\prime\prime}_{t,R}$.

\textbf{Initialization}: At time instant 1, the channel encoder maps the messages $W^{\prime\prime}_{t,R}$ to $\theta_{R}$, and sends
\begin{footnotesize}
\begin{align}\label{time1sends}
X_{R,1}=\sqrt{P_{R}}\theta_{R},
\end{align}
\end{footnotesize}%
at the end of time 1, the channel decoder of cloud server receives $Y_{1}$. Then, the decoder obtains $Y^{\prime}_{R,1}$
and computes the first estimation $\hat{\theta}_{R,1}$ of $\theta_{R}$ by
\begin{footnotesize}
\begin{align}\label{time1computes2}
\hat{\theta}_{R,1}=\frac{Y^{\prime}_{R,1}}{\sqrt{P_{R}}}=\theta_{R}+\frac{\eta^{\prime}_{R,1,1}}{\sqrt{P_{R}}}=\theta_{R}+\varepsilon_{R,1},
\end{align}
\end{footnotesize}%
where $\varepsilon_{R,1}=\hat{\theta}_{R,1}-\theta_{R}=\frac{\eta^{\prime}_{R,1,1}}{\sqrt{P_{R}}}$ is the decoding error of decoder at time instant 1. Define $\alpha_{R,1}=\text{Var}(\varepsilon_{R,1})=\frac{\sigma_{1}^{2}}{2|h|^{2}P_{R}}$.

\textbf{Iteration}: From the second time instant, we first introduce the dither signal sequence $V^{N_{t}-1}=(V_{1},...,V_{N_{t}-1})$, which follows from the extended SK-type feedback scheme in \cite{feedback}.
We assume that the i.i.d generated sequence $V^{N_{t}-1}$ is perfectly known by the edge server and the cloud server, where $V_{i}\sim \text{Unif}[-\frac{d}{2},\frac{d}{2}]$, $d=\sqrt{6\widetilde{P}}$.
  The dither signals ensure that the encoded codeword of cloud server satisfies the power constraint.

At time instant $i$ ($2\leq i \leq N_{t}$), the channel decoder of cloud server computes and sends
\begin{footnotesize}
\begin{align}\label{time2sends}
\widetilde{X}_{R,i-1}=\mathbb{M}_{d}[\gamma_{R,i-1}\hat{\theta}_{R,i-1}+V_{i-1}],
\end{align}
\end{footnotesize}%
where $\gamma_{R,i-1}$ is the modulation coefficient of cloud server, $\mathbb{M}_{d}$ is the modulo-$d$ function and it is defined in \cite{feedback}.
From property \romannumeral5\ of proposition 1 in \cite{feedback},  we have $E(\widetilde{X}^{2}_{R,i-1})=\frac{\widetilde{P}}{2}=\widetilde{P}_{R}$. After the channel encoder receives $\widetilde{Y}_{i-1}$, the channel encoder obtains $\widetilde{Y}^{\prime}_{R,i-1}$
and computes the noisy version of decoding error $\varepsilon_{R,i-1}=\hat{\theta}_{R,i-1}-\theta_{R}$ by
\begin{footnotesize}
\begin{align}\label{time2decerr1}
\widetilde{\varepsilon}_{R,i-1}&=\frac{1}{\gamma_{R,i-1}}\mathbb{M}_{d}[\widetilde{Y}^{\prime}_{R,i-1}-\gamma_{R,i-1}\theta_{R}-V_{i-1}]\nonumber\\
&\stackrel{(a)}=\frac{1}{\gamma_{R,i-1}}\mathbb{M}_{d}[\gamma_{R,i-1}\varepsilon_{R,i-1}+\eta^{\prime}_{R,2,i-1}],
\end{align}
\end{footnotesize}%
where (a) is due to property \romannumeral2\ of proposition 1 in \cite{feedback}. The \emph{modulo-aliasing errors} do not occur in channel encoder, if  suitable $\gamma_{R,i-1}$ is chosen such that $\gamma_{R,i-1}\varepsilon_{R,i-1}+\eta^{\prime}_{R,2,i-1}\in [-\frac{d}{2},\frac{d}{2})$.
Hence, the channel encoder obtains $\widetilde{\varepsilon}_{R,i-1}=\varepsilon_{R,i-1}+\frac{\eta^{\prime}_{R,2,i-1}}{\gamma_{R,i-1}}$.
Then, the channel encoder sends
\begin{footnotesize}
\begin{align}\label{time2sends2}
X_{R,i}=\lambda_{R,i-1}\gamma_{R,i-1}\widetilde{\varepsilon}_{R,i-1},
\end{align}
\end{footnotesize}%
where $\lambda_{R,i-1}$  is chosen to satisfy the input power constraints $P_{R}$ ($E[(X_{R,i})^{2}]=P_{R}$). Analogously, the channel decoder receives $Y_{i}$ and computes $Y^{\prime}_{R,i}$. Then,
the channel decoder updates $\hat{\theta}_{R,i}$ by computing
\begin{footnotesize}
\begin{align}\label{time2updates1}
\hat{\theta}_{R,i}=\hat{\theta}_{R,i-1}-\hat{\varepsilon}_{R,i-1}=\hat{\theta}_{R,i-1}-\beta_{R,i}Y^{\prime}_{R,i},
\end{align}
\end{footnotesize}%
where $\hat{\varepsilon}_{R,i-1}=\beta_{R,i}Y^{\prime}_{R,i}$, and $\beta_{R,i}=\frac{E(Y^{\prime}_{R,i}\varepsilon_{R,i-1})}{E(Y^{\prime}_{R,i})^{2}}$
is the Minimum Mean Square Error (MMSE) estimation coefficient, which ensures that $\varepsilon_{R,i-1}$ is  correctly estimated from $Y^{\prime}_{R,i}$.
Define $\varepsilon_{R,i}=\hat{\theta}_{R,i}-\theta_{R}$, (\ref{time2updates1}) yield that
\begin{footnotesize}
\begin{align}\label{time2decerr3}
\varepsilon_{R,i}=\varepsilon_{R,i-1}-\beta_{R,i}Y^{\prime}_{R,i},
\end{align}
\end{footnotesize}%
and define $\alpha_{R,i}=\text{Var}(\varepsilon_{R,i})$.

\textbf{Decoding}: At time instant $N_{t}$, the channel decoder  obtains the final estimation $\hat{\theta}_{R,N_{t}}=\theta_{R}+\varepsilon_{R,N_{t}}$. Then the channel decoder declares the center of sub-interval which $\hat{\theta}_{R,N_{t}}$  belongs to as the final estimation of the $\theta_{R}$. The channel decoder successfully decodes the message $W^{\prime\prime}_{t,R}$ if $\hat{\theta}_{R,N_{t}}$ is in the sub-interval of $\theta_{R}$, i.e., $\varepsilon_{R,N_{t}}\in[-\frac{\sqrt{3}}{2^{N_{t}R_{t,R}}},\frac{\sqrt{3}}{2^{N_{t}R_{t,R}}})$.

The following algorithm \ref{algo1} further explains the encoding-decoding scheme described above.
\begin{algorithm}
    \caption{ Encoding-decoding procedure of a sub-channel}
    \label{algo1}
    \begin{algorithmic}[1]
        \REQUIRE $W^{\prime\prime}_{t,R},\tau,N_{t},P_{R},\widetilde{P}_{R},\sigma_{1}^{2},\sigma_{2}^{2},h,\widetilde{h},V^{N_{t}-1},d=\sqrt{6\widetilde{P}}$
        \ENSURE $X_{R,N_{t}},\hat{\theta}_{R,N_{t}}$

       Initialization:

       Map $W^{\prime\prime}_{t,R}\rightarrow \theta_{R}$

       The edge server encodes $X_{R,1}=\sqrt{P_{R}}\theta_{R}$

       The cloud server computes $\hat{\theta}_{R,1}$ from (\ref{time1computes2})

       Iteration:
    \FOR  {$2\leq i\leq N_{t}$}
           \STATE Compute $\lambda_{R,i-1}$, $\gamma_{R,i-1}$, $\beta_{R,i}$ from (\ref{P2})-(\ref{parameters1})
           \STATE The cloud server encodes $\widetilde{X}_{R,i-1}$ from (\ref{time2sends})
           \STATE The edge server computes $\widetilde{\varepsilon}_{R,i-1}$  from (\ref{time2decerr1})
           \STATE The edge server encodes $X_{R,i}$ from (\ref{time2sends2})
           \STATE The cloud server computes $\hat{\theta}_{R,i}$ from (\ref{time2updates1})
    \ENDFOR
    \end{algorithmic}
\end{algorithm}

\subsection{Performance analysis}\label{secIII}

\subsubsection{Utility and privacy analysis\label{secIII-3-UP}}
First, note that since $\mathbf{W}_{t}$, $\boldsymbol\eta_{t}$ and $\mathbf{W}^{\prime}_{t}$ are i.i.d. generated, from Definition $1$, we conclude that
\begin{footnotesize}
\begin{eqnarray}\label{Ixy1}
\frac{1}{qT}\sum_{t=1}^{T}I(\mathbf{W}_{t};\mathbf{W}^{\prime}_{t})\leq\max_{t\in\{1,..,T\}}\frac{1}{2} \log \left(1+\frac{S_{\ell}\sigma_{w,t}^{2}}{K\sigma^{2}}\right)\leq \epsilon,
\end{eqnarray}
\end{footnotesize}On the other hand, from Definition $2$, we conclude that
\begin{footnotesize}
\begin{eqnarray}\label{constraint2}
\frac{1}{qT}\sum_{t=1}^{T}E(d(\mathbf{W}_{t},\mathbf{W}^{\prime}_{t}))=\frac{1}{qT}\sum_{t=1}^{T}E(||\boldsymbol\eta_{t}||^{2})=K\sigma^{2}\leq U.
\end{eqnarray}
\end{footnotesize}Combining (\ref{Ixy1}) and (\ref{constraint2}), (\ref{Th1PU}) in Theorem \ref{Th1} is proved.

\subsubsection{Parameter analysis\label{secIII-3-1}}
In our proposed scheme, we define the parameters $\lambda_{R,i}$, $\beta_{R,i}$, and $\gamma_{R,i}$  as follows, and the transmission performance of our scheme is determined by these parameters.
\begin{footnotesize}
\begin{gather}
\lambda_{R,i}=\sqrt{L\cdot\frac{P}{\widetilde{P}}},\ \ \gamma_{R,i}=\sqrt{\frac{1}{\alpha_{R,i}}\left(\frac{\widetilde{P}}{2L}-\frac{\sigma_{2}^{2}}{2|\widetilde{h}|^{2}}\right)},\label{P2}\\
\beta_{R,i}=\frac{\sqrt{2\alpha_{R,i-1}}}{\sigma_{1}}\frac{\sqrt{\text{SNR}(1-L\cdot\tilde{\text{SNR}}^{-1}|\widetilde{h}|^{-2})}}{\text{SNR}+|h|^{-2}},\label{P4}\\
\alpha_{R,i}=|h|^{-2}\text{SNR}^{-1}\left(1+\frac{\text{SNR} |h|^{2}}{\Psi_{1}\Psi_{2}}\right)^{1-i},\label{parameters1}
\end{gather}
\end{footnotesize}%
and $\Psi_{1}$, $\Psi_{2}$ and $L$   are defined in (\ref{parameters2-1})-(\ref{rng44}). Combining the above definitions of parameters, and through the error probability analysis, the achievable rate of this FBL scheme in $t$-th communication round is given by
\begin{footnotesize}
 \begin{align}\label{Rtotal}
 R_{t}=
 \frac{1}{N_{t}}\log\left(\frac{3 \text{SNR}|h|^{2}}{\left[Q^{-1}(\frac{\tau}{8})\right]^{2}} \left(1+\frac{\text{SNR} |h|^{2}}{\Psi_{1}\Psi_{2}}\right)^{N_{t}-1}\right).
 \end{align}
 \end{footnotesize}The parameters derivation and error probability analysis of the FBL scheme are similar to those in \cite{feedback}, hence we omit it here.

\subsubsection{Security analysis\label{secIII-3-1}}
First, we will perform a security analysis of our FBL scheme within the FBL regime, the eavesdropper's equivocation rate $\Delta$ can be re-written as
\begin{footnotesize}
\begin{align}\label{Equivocationfeedback1}
\Delta&=\frac{H(W^{\prime\prime}_{1},...,W^{\prime\prime}_{T}|Z^{N_{1}},...,Z^{N_{T}},h,\widetilde{h},g,\widetilde{g})}{H(W^{\prime\prime}_{1},...,W^{\prime\prime}_{T})}\nonumber\\
&\stackrel{(b)}=\frac{\sum_{t=1}^{T}H(W^{\prime\prime}_{t}|Z^{N_{t}},h,\widetilde{h},g,\widetilde{g})}{\sum_{t=1}^{T}H(W^{\prime\prime}_{t})},
\end{align}
\end{footnotesize}%
where (b) is due to the fact that $\mathbf{W}^{\prime}_{t}$ is mapped into $W^{\prime\prime}_{t}$ at each communication round, which indicates that $(W^{\prime\prime}_{1},...,W^{\prime\prime}_{T})$ and $(Z^{N_{1}},...,Z^{N_{T}})$ are independent of each other, and $H(W^{\prime\prime}_{t}|Z^{N_{t}},h,\widetilde{h},g,\widetilde{g})$ is given by
\begin{footnotesize}
\begin{align}\label{Equivocationfeedback2}
&H(W^{\prime\prime}_{t}|Z^{N_{t}},h,\widetilde{h},g,\widetilde{g})\nonumber\\
&\stackrel{(c)}\geq H(W^{\prime\prime}_{t}|gX_{1}(t)+\widetilde{g}\widetilde{X}_{1}(t)+\eta_{e,1}(t),...,gX_{N_{t}-1}(t)+\widetilde{g}\widetilde{X}_{N_{t}-1}(t)+\nonumber\\
&\eta_{e,N_{t}-1}(t),gX_{N_{t}}(t)+\eta_{e,N_{t}}(t),\eta_{1,1}(t),...,\eta_{1,N_{t}}(t),\eta_{2,1}(t),...,\eta_{2,N_{t}}(t),\nonumber\\
&\eta_{e,2}(t),...,\eta_{e,N_{t}}(t),\widetilde{X}_{1}(t),...,\widetilde{X}_{N_{t}-1}(t),h,\widetilde{h},g,\widetilde{g})\nonumber\\
&\stackrel{(d)}=H(W^{\prime\prime}_{t}|gX_{1}(t)+\eta_{e,1}(t),\eta_{1,1}(t),...,\eta_{1,N_{t}}(t),\eta_{2,1}(t),...,\eta_{2,N_{t}}(t),\nonumber\\
&\eta_{e,2}(t),...,\eta_{e,N_{t}}(t),\widetilde{X}_{1}(t),...,\widetilde{X}_{N_{t}-1}(t),h,\widetilde{h},g,\widetilde{g})\nonumber\\
&\stackrel{(e)}=H(W^{\prime\prime}_{t}|gX_{1}(t)+\eta_{e,1}(t))\nonumber\\
&\stackrel{(f)}=H(W^{\prime\prime}_{t})+h(\eta_{e,1}(t))-h(gX_{1}(t)+\eta_{e,1}(t))\nonumber\\
&\stackrel{(g)}=H(W^{\prime\prime}_{t})-\log(1+\frac{|g|^{2}P}{\sigma_{e}^{2}}),
\end{align}
\end{footnotesize}%
where
(c) follows from  (\ref{cioe}),
(d) follows from $X_{i}(t)=X_{R,i}(t)+jX_{I,i}(t)$ $(i=2,...,N_{t})$ is a function of $h,\widetilde{h},\eta_{1,1}(t),...,\eta_{1,i-1}(t),\eta_{2,1}(t),...,\eta_{2,i-1}(t)$,
(e) follows from the fact that $\widetilde{X}_{i}(t)=\widetilde{X}_{R,i}(t)+j\widetilde{X}_{I,i}(t)$ $(i=1,...,N_{t}-1)$ is only related to $V_{i}$ \cite[Chapter 4.1, pp. 61-63]{RZ}, and $h$, $\widetilde{h}$, $g$, $\widetilde{g}$, $\eta_{1,1}(t),...,\eta_{1,N_{t}}(t)$, $\eta_{2,1}(t),...,\eta_{2,N_{t}}(t)$, $\eta_{e,2}(t),...,\eta_{e,N_{t}}(t)$, $V_{1},...,V_{N_{t}-1}$ are independent of $W^{\prime\prime}_{t}$, $X_{1}(t)$, $\eta_{e,1}(t)$,
(f) is due to the fact that $X_{1}(t)=\sqrt{P_{R}}\theta_{R}+j\sqrt{P_{I}}\theta_{I}$, and $W^{\prime\prime}_{t}=(W^{\prime\prime}_{t,R},W^{\prime\prime}_{t,I})$ are mapped into $\theta_{R}$ and $\theta_{I}$, respectively, and
(g) is follows from
\begin{footnotesize}
\begin{align}\label{E7}
&h(gX_{1}(t)+\eta_{e,1}(t))-h(\eta_{e,1}(t))\nonumber\\
&\leq \log \text{det}\{\pi e[E(g(\sqrt{P_{R}}\theta_{R}+j\sqrt{P_{I}}\theta_{I})(\sqrt{P_{R}}\theta_{R}+j\sqrt{P_{I}}\theta_{I})^{\mathcal{H}}g^{\mathcal{H}})\nonumber\\
&+E(\eta_{e,1}(t)\eta_{e,1}(t)^{\mathcal{H}})]\}-\log \text{det}\{\pi e E(\eta_{e,1}(t)\eta_{e,1}(t)^{\mathcal{H}})\}\nonumber\\
&=\log(1+\frac{|g|^{2}P}{\sigma_{e}^{2}}).
\end{align}
\end{footnotesize}%
Substituting (\ref{Equivocationfeedback2}) into (\ref{Equivocationfeedback1}), we have
\begin{scriptsize}
\begin{align}\label{Equivocationfeedbackbbb}
\Delta&\geq\frac{\sum_{t=1}^{T}H(W^{\prime\prime}_{t})(1-\frac{\log(1+\frac{|g|^{2}P}{\sigma_{e}^{2}})}{H(W^{\prime\prime}_{t})})}{\sum_{t=1}^{T}H(W^{\prime\prime}_{t})}\geq\min_{t\in\{1,...,T\}}(1-\frac{\log(1+\frac{|g|^{2}P}{\sigma_{e}^{2}})}{H(W^{\prime\prime}_{t})}).
\end{align}
\end{scriptsize}%
From  (\ref{qR}) and (\ref{Equivocationfeedbackbbb}), $\Delta\geq\delta$ in (\ref{Def4}) is guaranteed if
\begin{footnotesize}
\begin{align}\label{fblN1}
\delta\leq\min_{t\in\{1,...,T\}}[1-\frac{\log(1+\frac{|g|^{2}P}{\sigma_{e}^{2}})}{qR_{t}(D)}]^{+}.
\end{align}
\end{footnotesize}%
Then, the secrecy achievable rate $R(N,\tau,U, \epsilon,\delta,D)$ is given by
\begin{scriptsize}
\begin{align}\label{TotR}
R(N,\tau,U, \epsilon,\delta,D)=\frac{H(W^{\prime\prime}_{1},...,W^{\prime\prime}_{T})}{N}\stackrel{(h)}=\frac{\sum_{t=1}^{T}H(W^{\prime\prime}_{t})}{N}=\frac{\sum_{t=1}^{T}N_{t}R_{t}}{N},
\end{align}
\end{scriptsize}%
where (h) is similar to (b).
Finally, combining the (\ref{Ixy1}), (\ref{constraint2}), (\ref{Rtotal}), (\ref{fblN1}) and (\ref{TotR}), we complete the proof of Theorem \ref{Th1}.

\section{ Simulation Results\label{secIV}}
\setcounter{equation}{0}
Similar to \cite{GZB} and \cite{b}, we assume that the channel coefficients are i.i.d. as $\mathcal{CN}(0,1)$, here note simulation results are based on an average of 1000 independent channel realizations.
We consider a wireless HFL system with $K=10$ users, a edge server and a cloud server, and the training samples are uniformly distributed across the 10 users. The regularization function $R(\mathbf{m})=||\mathbf{m}||^{2}$ with $\lambda=5\times10^{-5}$. Before the channel coding, the edge server compresses the quantified data via Lempel Ziv Welch (LZW) source coding \cite{LZW}, and the total data amount to be transmitted is defined as $M$ bits. The wireless data transmission latency for uploading information of edge server can be calculated by $T_{c}=M/R_{eg}$ \cite{GZB}, where $R_{eg}$ represents the transmission rate of edge server.
We train a neural network on the MNIST data set\footnote{http://yann.lecun.com/exdb/mnist/}, and the neural network consists of 784 input nodes, a single hidden layer with 20 hidden nodes, and 10 output nodes. We use the cross entropy as the loss function, and the rectified
linear unit (ReLU) and the softmax functions are the activation functions of the hidden and output layers, respectively.
The total number of parameters in the neural network is $q=15910$ and the learning rate be $\mu=1$.

\begin{figure}[htbp]
    \centering
    \subfigure[Test accuracy]{
    \begin{minipage}[t]{0.48\linewidth}
    \centering
    \includegraphics[scale=0.31]{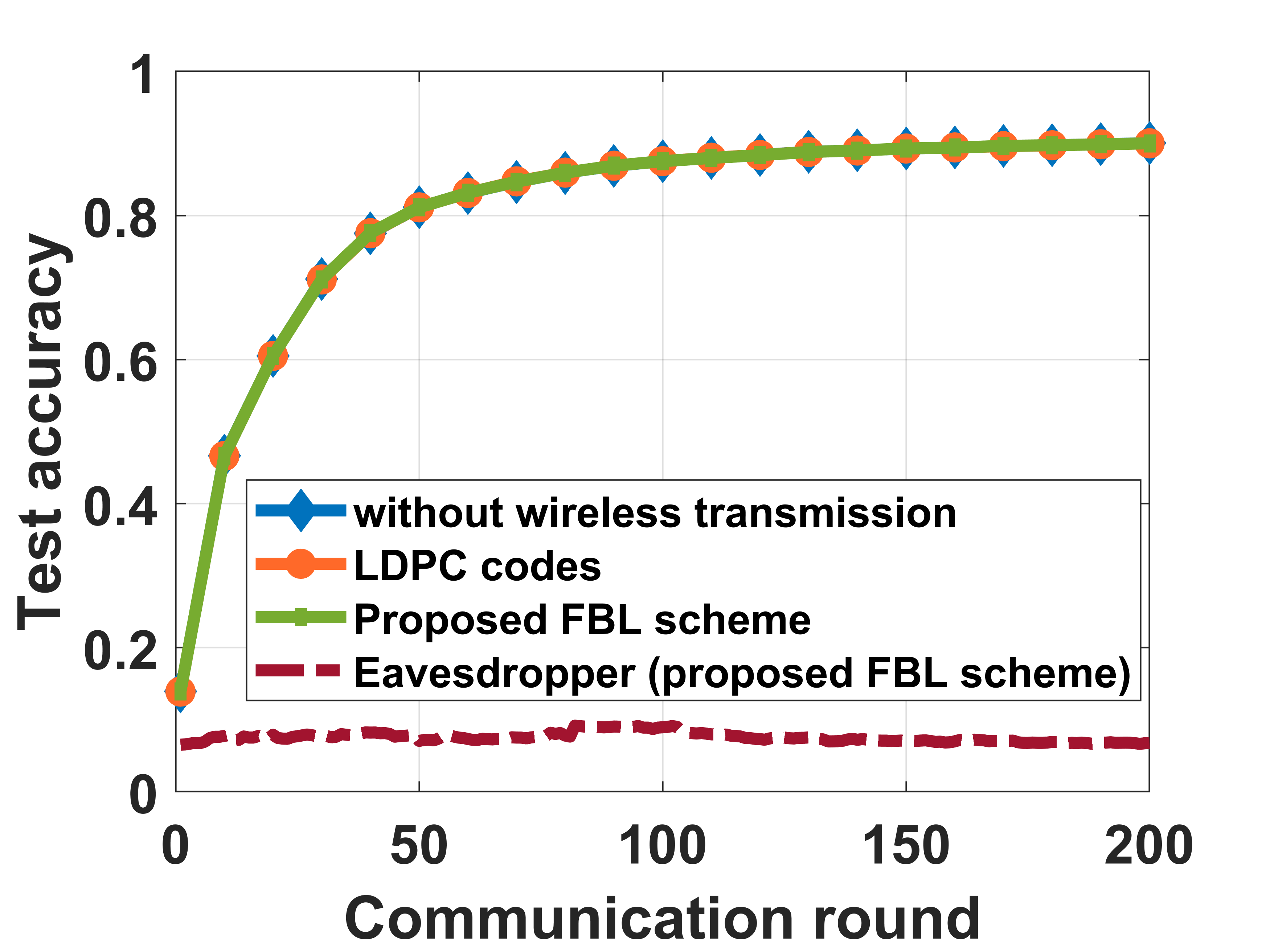}\label{accuracy}
    \end{minipage}%
    }%
    \subfigure[Training loss]{
    \begin{minipage}[t]{0.48\linewidth}
    \centering
    \includegraphics[scale=0.31]{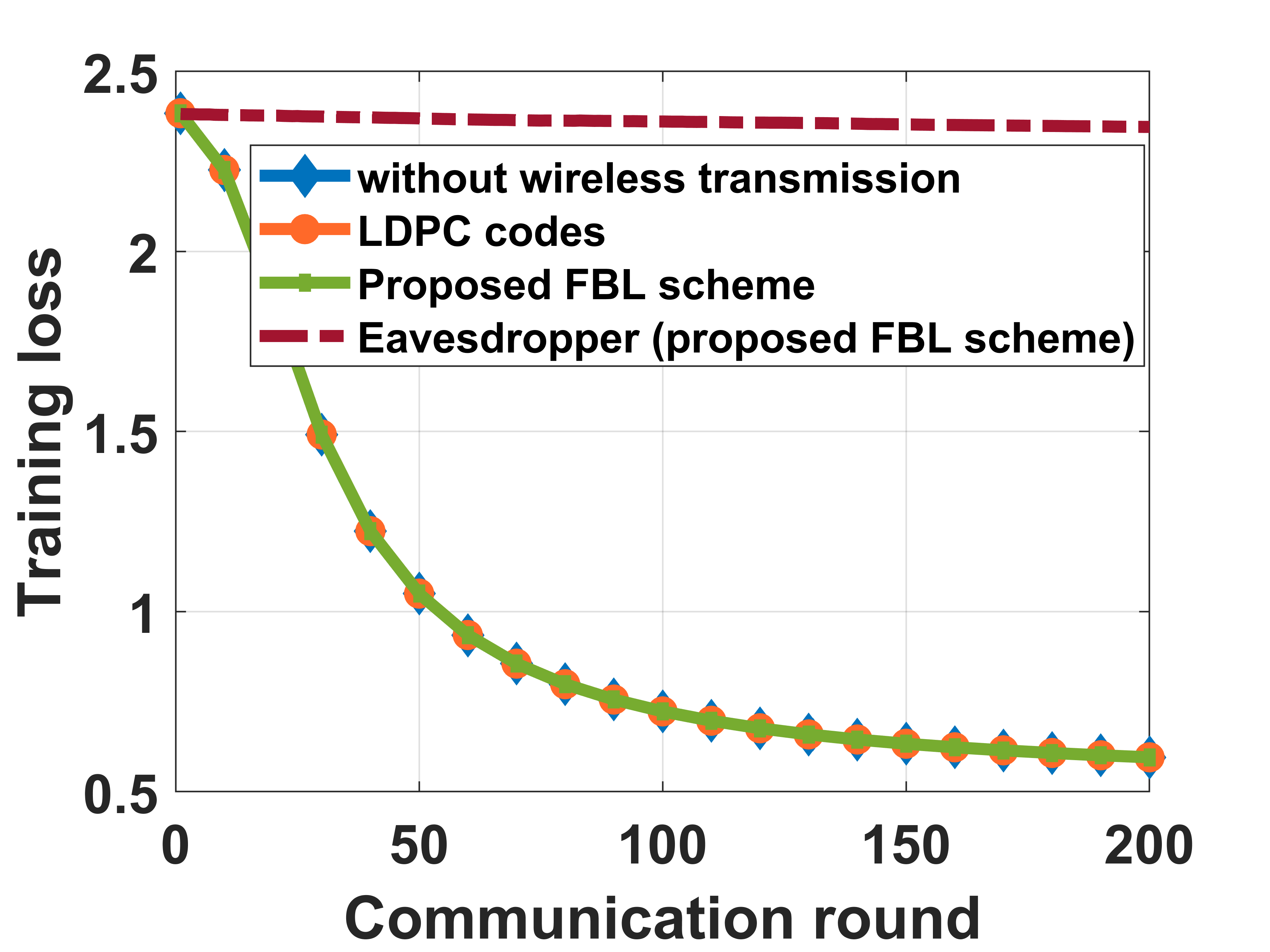}\label{lossnn}
    \end{minipage}%
    }%
    \caption{ Performance comparison between the different schemes
  ($U=5$, $\epsilon=0.1$, $D=10^{-4}$, $\tilde{\text{SNR}}=15$dB, $\tau=10^{-6}$, $\sigma^{2}=0.5$, $\sigma_{1}^{2}=\sigma_{2}^{2}=\sigma_{e}^{2}=1$, $P=10$, $S_{\ell}=60000$)}
    \label{MNIST1}
\end{figure}

\begin{figure}[htb]
\centering
\includegraphics[scale=0.32]{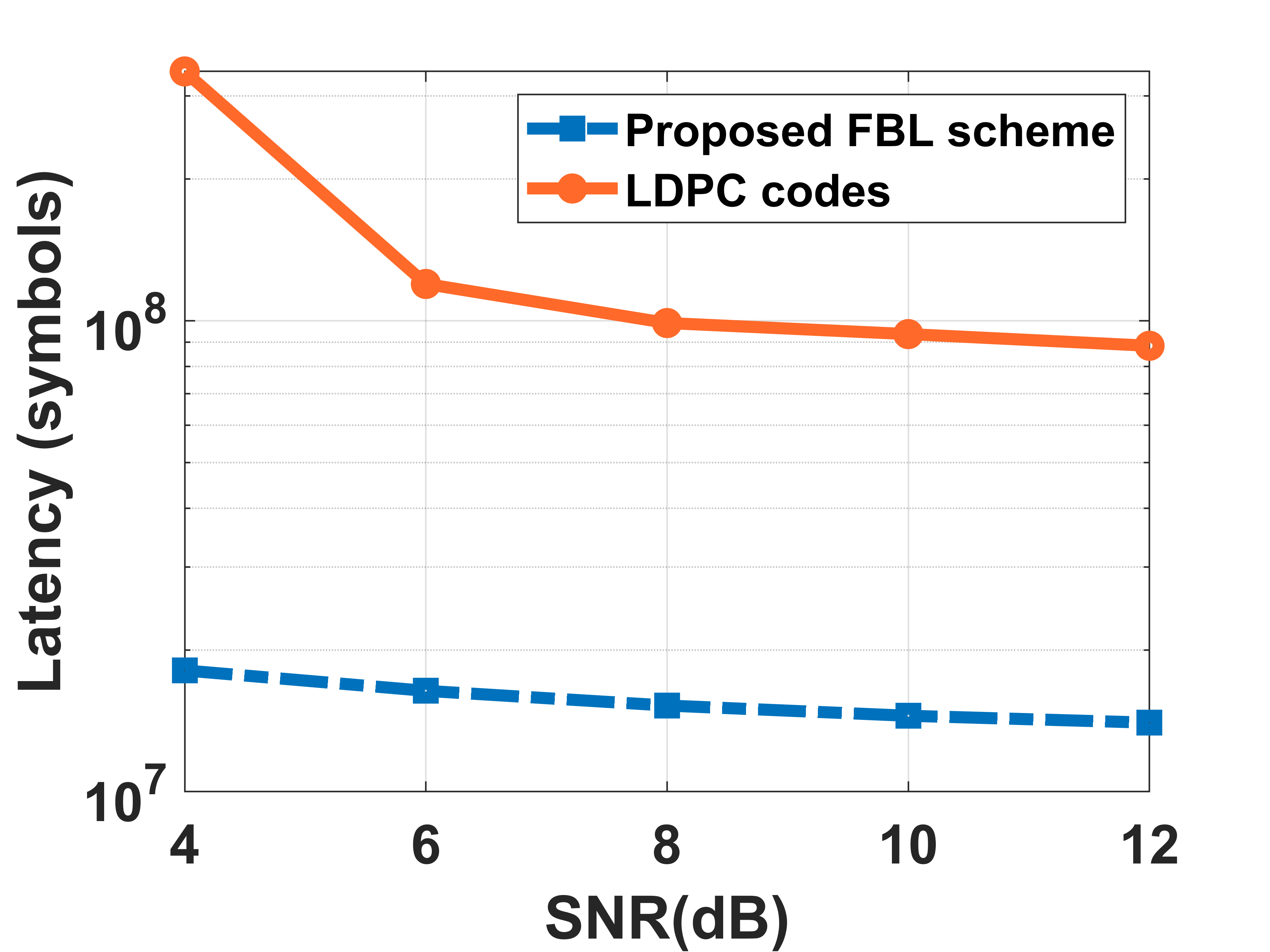}
\caption{Transmission latency  with different schemes ($U=5$, $\epsilon=0.1$, $D=10^{-4}$, $\tilde{\text{SNR}}=15$dB, $\tau=10^{-6}$, $\sigma^{2}=0.5$, $\sigma_{1}^{2}=\sigma_{2}^{2}=\sigma_{e}^{2}=1$, $T=200$, $S_{\ell}=60000$)}
\label{figlat}
\end{figure}

\begin{figure}[htbp]
    \centering
    \subfigure[Test accuracy]{
    \begin{minipage}[t]{0.48\linewidth}
    \centering
    \includegraphics[scale=0.115]{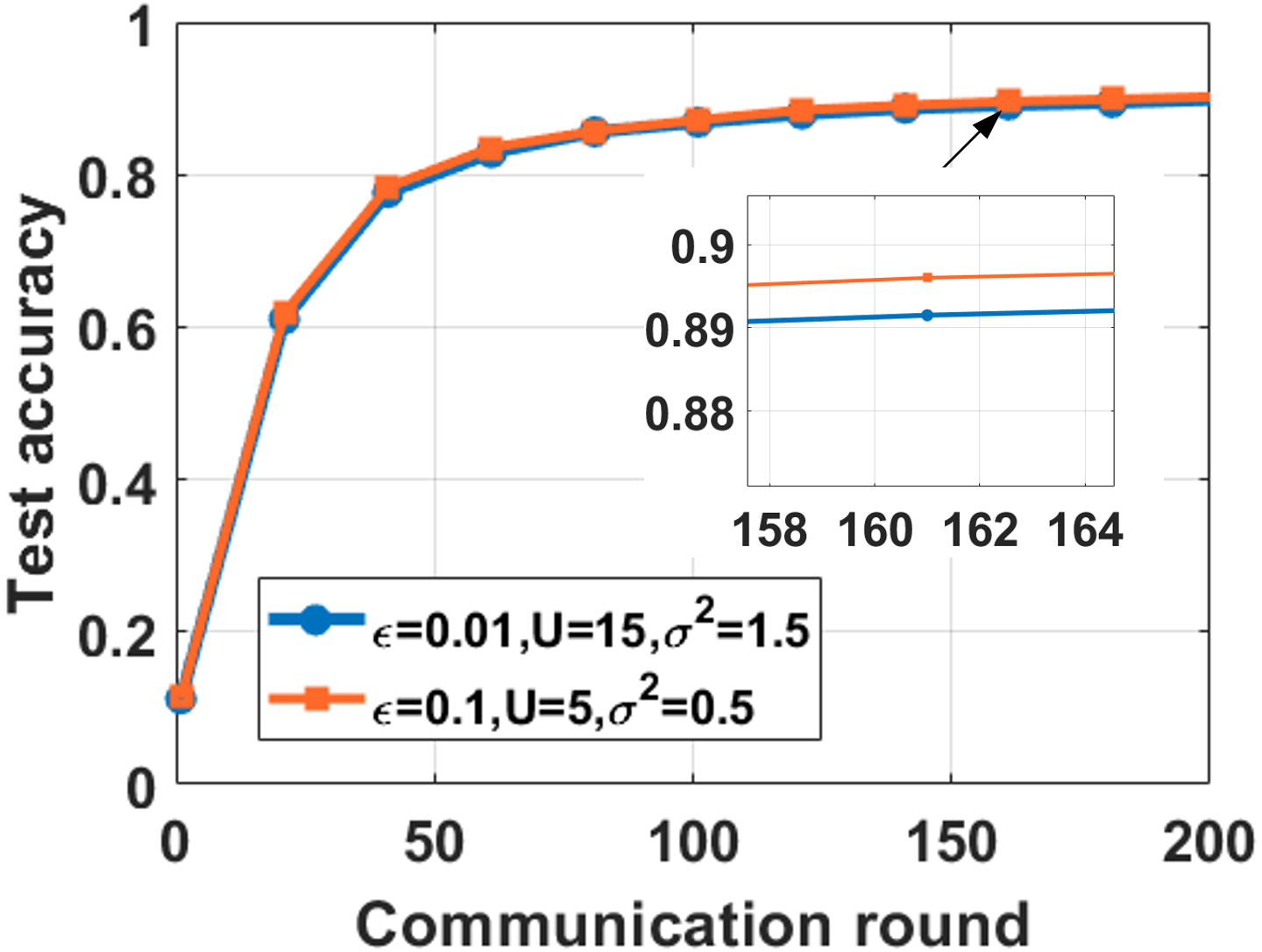}\label{accuracyup}
    \end{minipage}%
    }%
    \subfigure[Training loss]{
    \begin{minipage}[t]{0.48\linewidth}
    \centering
    \includegraphics[scale=0.31]{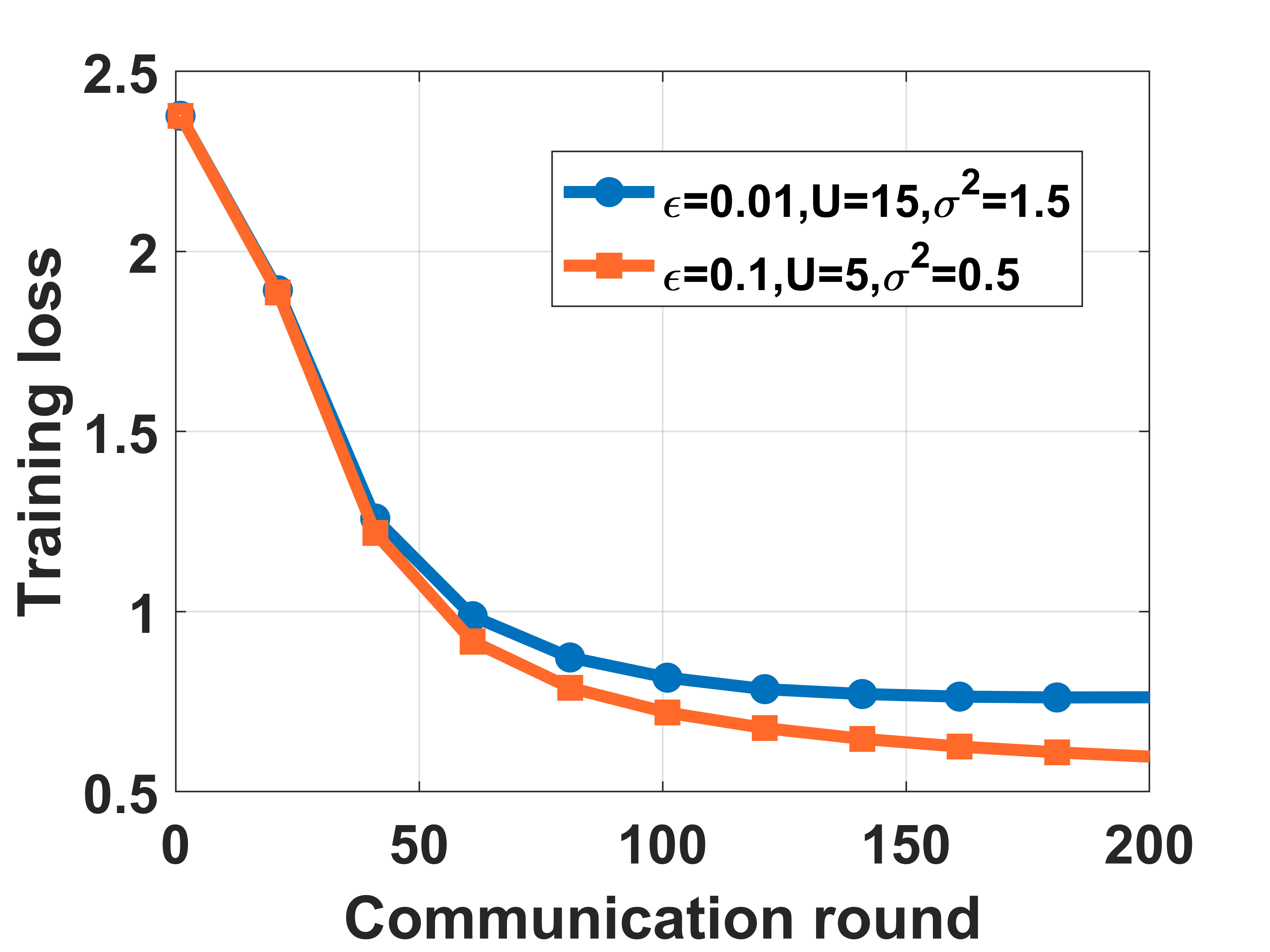}\label{lossnnup}
    \end{minipage}%
    }%
    \caption{ Performance comparison between the different $U$ and $\epsilon$ of proposed FBL scheme
  ($\tilde{\text{SNR}}=15$dB, $\tau=10^{-6}$, $\sigma_{1}^{2}=\sigma_{2}^{2}=\sigma_{e}^{2}=1$, $P=10$, $D=10^{-4}$, $S_{\ell}=60000$)}
    \label{MNIST2}
\end{figure}

\begin{figure}[h]
    \centering
    \subfigure[Achievable secrecy rates ]{
    \begin{minipage}[t]{0.48\linewidth}
    \centering
    \includegraphics[scale=0.31]{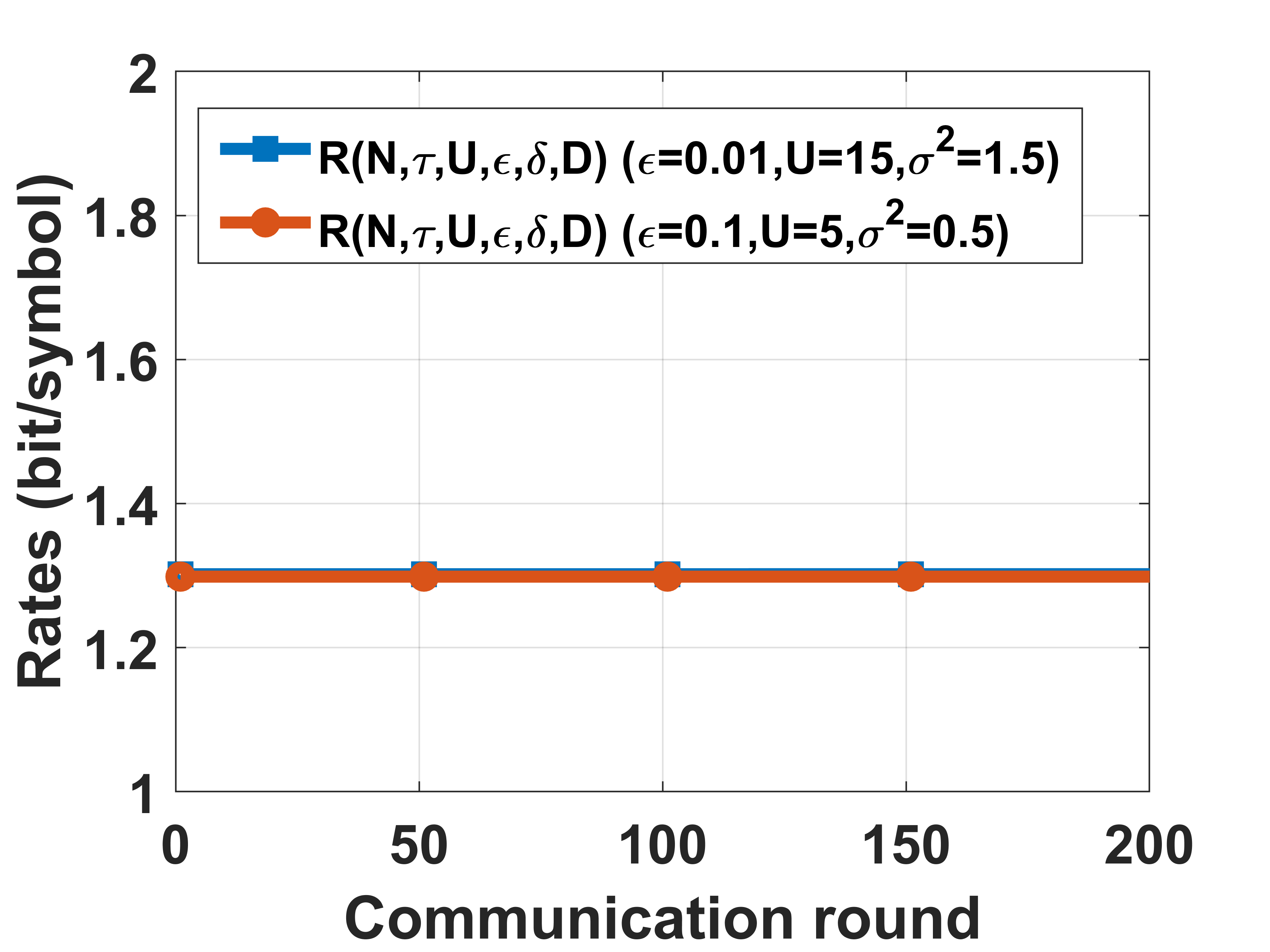}\label{Rtot}
    \end{minipage}%
    }%
    \subfigure[Secrecy level]{
    \begin{minipage}[t]{0.48\linewidth}
    \centering
    \includegraphics[scale=0.115]{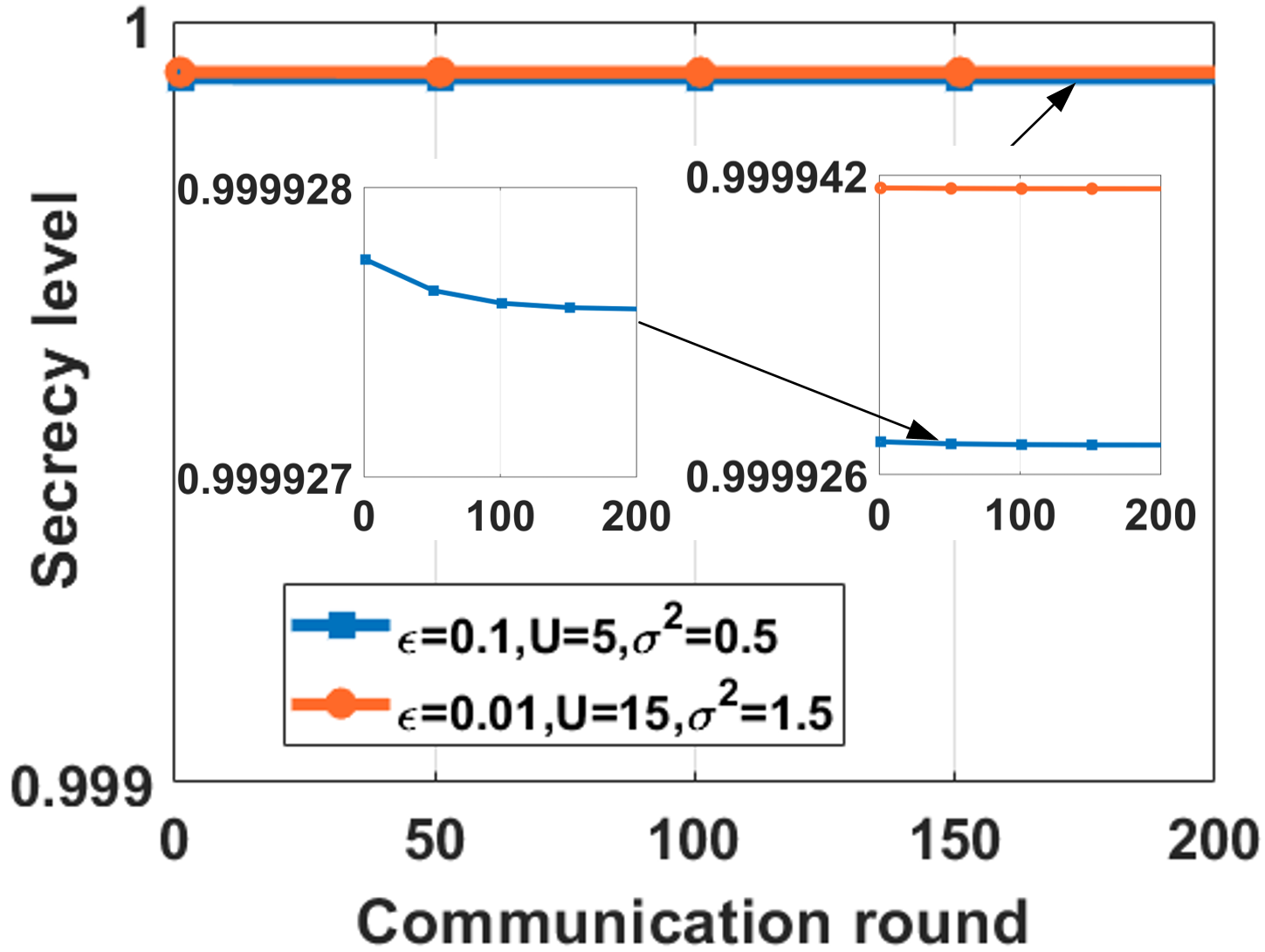}\label{level}
    \end{minipage}%
    }%
    \caption{ Performance of proposed FBL scheme
  ($\tilde{\text{SNR}}=15$dB, $\tau=10^{-6}$, $\sigma_{1}^{2}=\sigma_{2}^{2}=\sigma_{e}^{2}=1$, $P=10$, $D=10^{-4}$, $S_{\ell}=60000$)}
    \label{MNIST3}
\end{figure}

From Figure \ref{accuracy} and Figure \ref{lossnn}, we see that the channel coding does not affect the learning performance of HFL, and the eavesdropper has a poor learning performance when using our proposed FBL scheme, which indicates that the PLS of the data is guaranteed by the proposed FBL scheme. In Figure \ref{figlat}, we conclude that the transmission latency of the proposed FBL scheme is significantly low compared with LDPC codes (10x less latency).
Figure \ref{MNIST2} plots the learning performance of our FBL scheme under different privacy-utility constraints, we conclude that more stringent privacy-utility constraints (smaller $\epsilon$ and larger $U$) lead to lower learning performance. The different privacy-utility constraints do not affect the achievable secrecy rates, as shown in Figure \ref{Rtot}, because the achievable secrecy rate approaches a constant as blocklength increases.
Figure \ref{level} shows that the secrecy level increases as the privacy-utility constraints become more stringent. Moreover, as the communication round increases, the secrecy level decreases because the variance of the gradient decreases as the training continues \cite{ZJ}.

\section{Conclusion and Future Work\label{sec5}}
This paper proposes a practical FBL coding scheme for the wireless HFL in the presence of PLS, which almost achieves perfect secrecy without affecting learning performance. Besides this, simulation results show that
the coding blocklength of our proposed scheme is significantly shorter than classical LDPC code.
One possible future work is to study the case that imperfect CSI is obtained by all parties.

\end{document}